\documentclass{JINST}
\usepackage{cite}
\usepackage{graphicx}
\usepackage{amssymb}
\usepackage{amsthm}
\usepackage{amstext}
\usepackage{epsfig}
\usepackage[english]{babel}

\usepackage{lineno}

\graphicspath{{./figures/}}

\title{Eco-friendly gas mixtures for Resistive Plate Chambers based 
on Tetrafluoropropene and Helium}

\author{
  M.~Abbrescia$^{1}$,
  L.~Benussi$^{2}$,
   S.~Bianco$^{2}$,
  M.~Ferrini$^{3}$,
  S.~Muhammad$^{3}$,
  L.~Passamonti$^{2}$,
  D.~Pierluigi$^{2}$,
  D.~Piccolo$^{2}$,
  F.~Primavera$^{2}$,
  A.~Russo$^{2}$,
  G.~Saviano$^{3}$\\
  \llap{$^1$} INFN and Dipartimento di Fisica, Universit\`a di Bari, Bari, Italy\\
  \llap{$^2$} INFN Laboratori Nazionali di Frascati, Frascati (RM), Italy\\
  \llap{$^3$} INFN Laboratori Nazionali di Frascati and Facolt\'a di Ingegneria Roma1 (RM), Italy\\

  E-mail: \email{marcello.abbrescia@ba.infn.it, davide.piccolo@lnf.infn.it}
}
\date{Received: date / Revised version: date}
\abstract{
Due to the recent restrictions deriving from the application of the Kyoto protocol,
the main components of the gas mixtures presently used in the Resistive Plate Chambers systems of
the LHC experiments will be most probably phased out of production in the coming years. 

Identifying possible replacements with the adequate 
characteristics requires an intense R\&D,
which was recently started, also in collaborations across the various
experiments. Possible candidates have been proposed and are thoroughly investigated.

Some tests on one of the most promising candidate - HFO-1234ze, an allotropic form 
of tetrafluoropropane- have already been reported. Here an innovative approach, based 
on the use of Helium,
to solve the problems related to the too elevate 
operating voltage of HFO-1234ze based gas mixtures, 
is discussed and the relative first results are shown.
}

\keywords{tetrafluoroethane, tetrafluoropropane, helium, eco-friendly gas mixtures, Global Warming Power}

\begin{document}

\section{Introduction}

Resistive Plate Chambers (RPCs in the following) were invented at the beginning 
of 1980s and originally thought to be
operated in streamer mode, using Argon based gas mixtures~\cite{Santonico}.
Later on, when they were proposed for the experiments at Large Hadron Collider, 
the rate limitation
intrinsic to operate these devices in streamer mode was overcome by passing
to some kind of "saturated" avalanche mode, obtained by filling them with Freon
based gas mixtures and transferring part of the needed amplification from 
the gas to the front-end electronics ~\cite{Cardarelli}~\cite{Camarri}.

The first Freon used in RPC, CF$_3$Br (at first added just in small percentages to 
the Argon based mixtures, and later on used as the main component in the first attempts 
towards the avalanche
mode) was soon prohibited due to regulations about environment protection, 
since it was known to damage the ozone atmospheric layer.

Tetrafluoroethane C$_2$H$_2$F$_4$ (commercially known as R-1234a), 
characterized by a negligible ozone depletion
power, was introduced to replace it, and soon adopted by all LHC experiments 
as the main component of their RPC gas mixtures. It was successfully used 
throughout all R\&D tests, the experiments commissioning phase, Run-1 and
it is currently used during Run-2. RPCs filled with it contributed to the Higgs
boson and other important discoveries.

Nevertheless, C$_2$H$_2$F$_4$ is characterized by a Global Warming Power (GWP) around 1430
(the reference being the one of carbon dioxide 
GWP(CO$_2$)=1), and recent regulations from the European Community
derived from the adoption of the Kyoto protocol, prohibit for many applications the use of gas mixtures
with a GWP>150. Even if scientific laboratories are
explicitly excluded from this prohibition, CERN is pushing the LHC experiments collaborations
to investigate for possible replacements of C$_2$H$_2$F$_4$ with other more eco-friendly
gases. 

This has led to many R\&D studies, sometimes coordinated across the 
different experiments. This was necessary, since finding a new RPC gas mixture is quite
complicated, due to the many possible candidates, many possible combinations and 
different percentages that could in principle be used and need to be tested.

One of the possible ideas to follow, proposed some time ago by one of the authors 
of this paper~\cite{Private}, relies on finding the molecule
most similar to C$_2$H$_4$F$_4$ characterized by a low enough GWP. 
Of course, this is not
sufficient to assure that the new gas would be suitable for RPC operation, and
the corresponding tests to assure that RPCs filled with this mixture provide the
necessary efficiency, time resolution, rate capability, aging tolerance, are
anyhow needed. On the other hand, the criteria proposed above seem to be quite
reasonable to make educated guesses about possible candidates in the huge amount
of possibilities.

A molecule quite similar to C$_2$H$_2$F$_4$ but characterized with a GWP<150 is 
tetrafluoropropane, C$_3$H$_2$F$_4$; the two molecules, essentially, differ by a one single
atom (one carbon atom more with respect to the tetrafluorethane). 
Tetrafluoropropane also comes in two allotropic forms, commercially indicated 
as HFO-1234yf and HFO-1234ze.
Both gases satisfy the requirement about their GWP, being GWP(HFO-1234yf)=4,
and GWP(HFO-1234ze)=6. 

However, one of the two, HFO-1234yf, is reported to be mildly flammable,
and this induces to start testing the other.
Both of them are quite expensive at the moment (around ten times the cost of
R-1234a), but it must be stressed that
R-1234a is extensively used as refrigerator, for instance in automobiles refrigerating plants; 
once that it will be phased out,
HFO-1234ze is one of the most interesting candidates to replace it, 
and this would let us suppose that its price
will be progressively decreasing in the coming years.

Tests on mixtures obtained replacing increasing fractions of tetrafluoroetane with
tetrafluoropropane have already been performed, and are still in progress 
~\cite{Frascati}~\cite{Liberti}.
Basically, they demonstrate that HFO-1234ze based gas mixtures could be a promising solution
validating the proposal made above, but with the
main drawback that the operating voltage is shifted toward much higher
values with respect to tetrafluoroethane based gas mixtures.
2 mm gap RPCs of the type used in CMS and ATLAS filled with HFO gas mixtures, for instance, 
would typically operate at operating voltage much higher than 10 kV, the exact value
depending on the precise mixture composition.
This is related to the fact that HFO-1234ze has an effective first Townsend coefficient
lower, at the same electric field, with respect to tetrafluroethane. 

In principle, this would not be a problem. However, the present RPC systems
of the LHC experiments are designed to operate at voltages lower than 10 kV
(in particular in terms of high voltage cables and connectors), therefore it would
be interesting to identify a gas mixture with the requirement -in addition
to the ones already cited- that the
present RPCs could operate at voltages lower than 10 kV.
This could be obtained adding Helium to HFO-1234ze based gas mixtures.

Helium is a gas characterized by no vibrational and rotational degrees of freedom,
and with a large first ionization energy, being 
around 24.6 eV, much higher than tetrafluoroethane or tetrafluoropropane. This has
the consequence that, in a mixture composed mainly by tetrafluoroethane or tetrafluoropropane, 
most ionising interactions take place on these two gases
and not on Helium, and that 
when the average energy of the 
electron avalanche is not large enough, most electron interactions with 
Helium atoms are essentially elastic.  
In this approximation
Helium does not take any part in the avalanche process, but behaves just as a "space holder" in the 
gas volume; as a consequence the partial
pressure of the "active" gas mixture, in this case composed essentially 
by tetrafluoroethane or tetrafluoropropane, is reduced by an amount corresponding
to the fraction of Helium in the mixture.

It is well know that a reduction in the pressure of the gas mixture has the 
consequence to reduce also the operating voltage
of RPCs (or other gaseous detectors), since it increases the average mean free path, 
and consequently the first effective Townsend coefficient. Roughly, gas parameters 
are the same for the same value of
the $E/d$ ratio, where $E$ is the electric field strength and $d$ in the gas density
~\cite{Pressione1}~\cite{Pressione2}.

So, in presence of gas mixtures characterized by a high operating 
voltage, it is a natural idea to add a fraction of Helium
to reduce their partial pressure and lower the operating voltage 
while the performance keeps stable. 
This has already been proved some years ago ~\cite{Helium}, 
with quite encouraging results.

In this paper, the first results of tests using HFO-124ze based gas 
mixtures, to which some fraction of Helium has been added, will be reported.

\section{Experimental set-up}

This work was performed using the very same experimental set-up already 
employed to obtain the results described in~\cite{Frascati}; therefore just its main 
characteristics will be reported here.
It consists of twelve bakelite 50 $\times$ 50 cm$^2$ single-gap Resistive Plate 
Chambers located at the Frascati National Laboratories of the Italian Institute for 
Nuclear Physics, and arranged on top of each other in a telescope for cosmic rays.  

Each chamber is read-out by means of two copper pads glued on the opposite external 
sides of the electrode plates. The pads are read out by a custom circuit which algebraically 
subtracts the signals coming from the two gaps; since these are of opposite polarities, this eliminates the 
coherent noise and adds up the signals induced by the charges moving inside the gas gap, 
increasing significantly the signal to noise ratio.

The trigger is provided by two scintillators located on the top and bottom of the RPC stack; given 
the measured 1 Hz average trigger rate, 1000 events are collected in about 20 minutes. 
Signals from the pads are digitalized by means of a Lecroy oscilloscope, with a 1 GHz bandwidth; 
for each signals 5 Gsamples are taken per event, covering a time span of about 1 $\mu$s.
For these tests, just two RPCs were read out, both flushed with the gas 
mixtures under study. One of them, however, 
was kept at fixed voltage and used as an additional off-line trigger, 
to furtherly reduce trigger accidentals.

For the RPC under test, measurements were taken at different effective High
Voltages (henceforth labelled as $HV_{eff}$), using the standard formula
$HV_{eff} = HV_{appl} \times p_{0}/p \times T/T_{0}$ derived from~\cite{Pressione1} to
relate it to the applied voltage $HV_{appl}$, where, 
in this case using
$p_{0}$ = 990 mbar and $T_{0}$= 293.15 K have been used 
as reference pressure and temperature respectively.
 
In order to reduce the environmental noise picked up from the system, a 200 MHz 
filter was connected in input to the oscilloscope; this, in addition to 
using pad read-out electrodes, has the effect to increase the signal rise time, slightly worsening 
the measured time resolution; the measured induced charge, on the contrary, being
an integral quantity, should not be affected.

For analysis purposes, waveforms acquired by means of the oscilloscope and 
stored, are divided into four "regions". The region of 200 ns centerd around 
the trigger is the "signal region", used to measure signal characteristics. The 50 ns 
region before it, where signals are absent, is the "baseline region", and it is used to 
measure the average baseline to be subtracted to the measured values taken in the "signal region". 
In addition, the 200 ns before the "baseline window" is called the 
"control region" and is used to check the width of the charge distribution 
and of the peak voltage, in absence of signal, to define the correct threshold to identify RPC signals. 
The 200 ns after the "signal region" is called "delayed region" and is used to search 
for signals delayed with respect to the main one that could be related to streamers.

Induced charge is computed by integrating the waveform -after 
baseline subtraction- in the "signal window", and dividing the result 
by the oscilloscope input impedance. 
To distinguish RPC signals from noise, it is required that they 
are characterized by an integrated charge greater than 300 fC, 
which is quite close to the usual thresholds used for the RPC 
in CMS and ATLAS. In addition the signals are required to be higher than a minimum 
voltage of 0.4 mV, in absolute value.

A streamer is defined as a signal characterized by an integrated 
charge larger than 40 pC in the 400 ns time window defined by the 
"signal" plus "delayed" regions. Finally the time resolution is obtained by 
measuring the time the signals from the two RPCs cross over a threshold of -0.8 mV 
and plotting the difference. The time resolution of the trigger RPC is deduced looking
at the time difference distribution taken when both trigger and test RPCs have the same effective voltage.  
Assuming that at the same voltage both RPCs have the same resolution, 
the time resolution of the trigger RPC is
deduced dividing by $\sqrt{2}$ the sigma of the difference distribution. 
Finally the time resolution of the test RPC at different applied voltages is extracted subtracting in
 quadrature the trigger RPC resolution to the measured sigma of the time differences at each voltage.

\section{Results from Helium and Tetrafluoropropane based gas mixtures}

\begin{figure}[tbp]
  \begin{center}
    \includegraphics[width=0.8\textwidth]{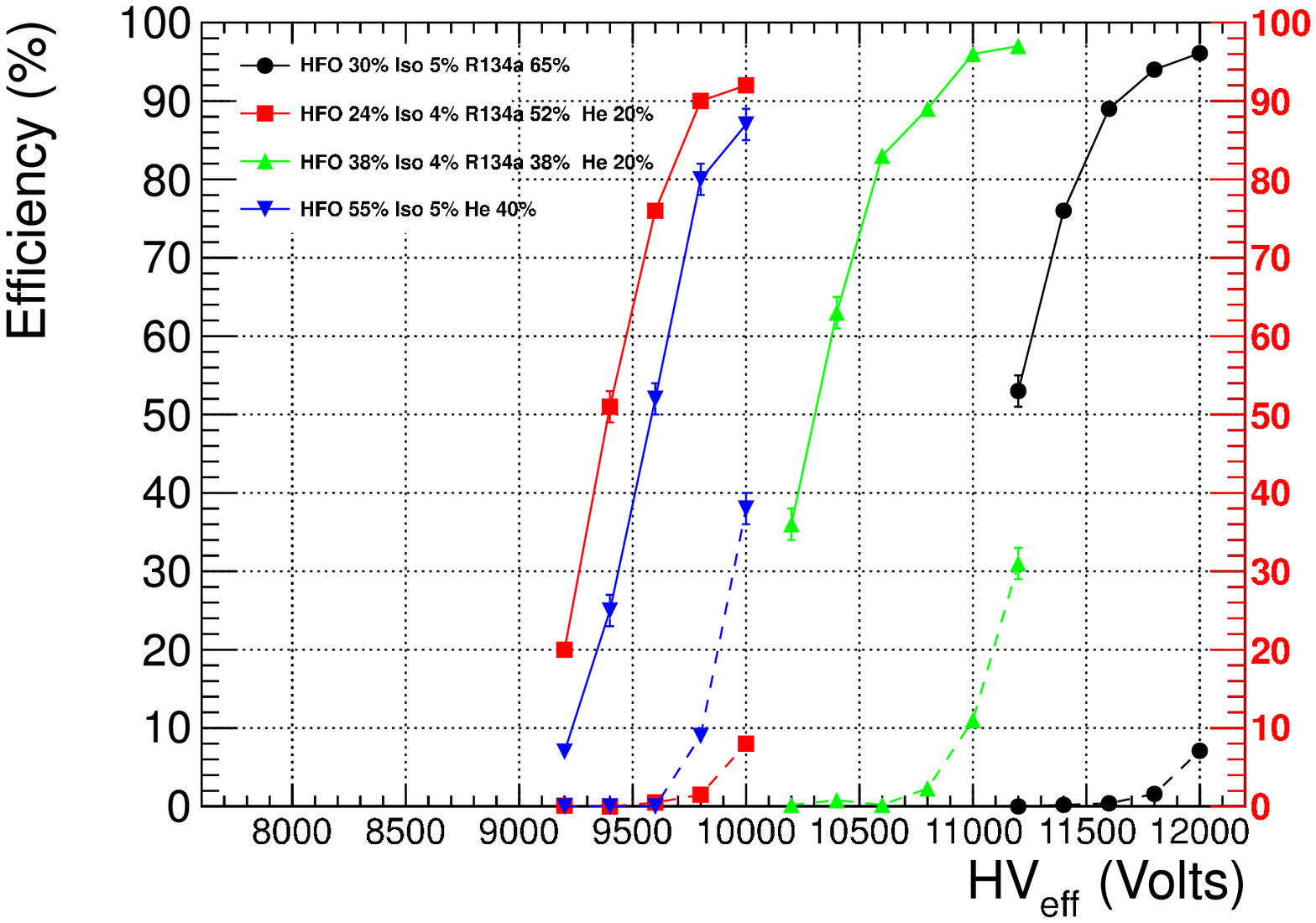}
    \caption{Efficiency (solid line) and streamer probability
(dashed line) as a function of effective voltage $HV_{eff}$, for
tetrafluoropropene/iso-buthane/Helium based gas mixtures}
    \label{fig:eff1}
  \end{center}
\end{figure}

In this section the first results of tests using HFO based gas mixtures to which
some fraction of Helium has been added are reported.
In order to validate the Helium power to reduce the working voltage, 
we started from a gas mixture made from 
65 $\%$ of tetrafluoroethane, 30 $\%$ of tetrafluoropropane and 5  $\%$ of iso-buthane. 
With this gas mixture the working voltage is around 12 kV. Note that this 
gas mixture does not fullfill the requirement GWP<150
(essentially because of the presence of C$_2$H$_2$F$_4$, nevertheless
it is useful to understand the role of the various gases
with respect to the operating voltae. To this gas
mixture increasing fractions of Helium were added; also the ratio between 
tetrafluroethane and tetrafluoropropane was modified. 

The effects of adding Helium is 
clearly visible in figure ~\ref{fig:eff1}. 
The efficiency (in solid line) and the streamer probability 
(in dashed line) are shown in the plot as a function of the effective voltage $HV_{eff}$, as defined
in the previous section. 
The black line refers to the gas mixture used as a starting point, 
composed by 65 $\%$ of R-134a, 30 $\%$ of HFO-1234ze and 5 $\%$ of iso-buthane; 
the red line refers to the same gas mixture, but with
the addition of 20 $\%$ of Helium, while keeping constant the ratios among R134a, 
HFO-1234ze and iso-buthane.
Adding 20 $\%$ of Helium has the effect to shift the efficiency curves toward
lower values of $HV_{eff}$, of an amount of around 2 kV, as expected.
 
In the same plot, the green line shows the effect of increasing the relative fraction of HFO-1234ze 
(and correspondigly reducing the one of R-134a), 
while keeping Helium and iso-buthane to 20 and 4 $\%$ respectively.
Finally the Helium fraction was increased to 40 $\%$, while keeping iso-buthane at 
5 $\%$ and completed the mixture with HFO-1234ze, and this is shown by the blue line.

\begin{figure}[tbp]
  \begin{center}
    \includegraphics[width=0.8\textwidth]{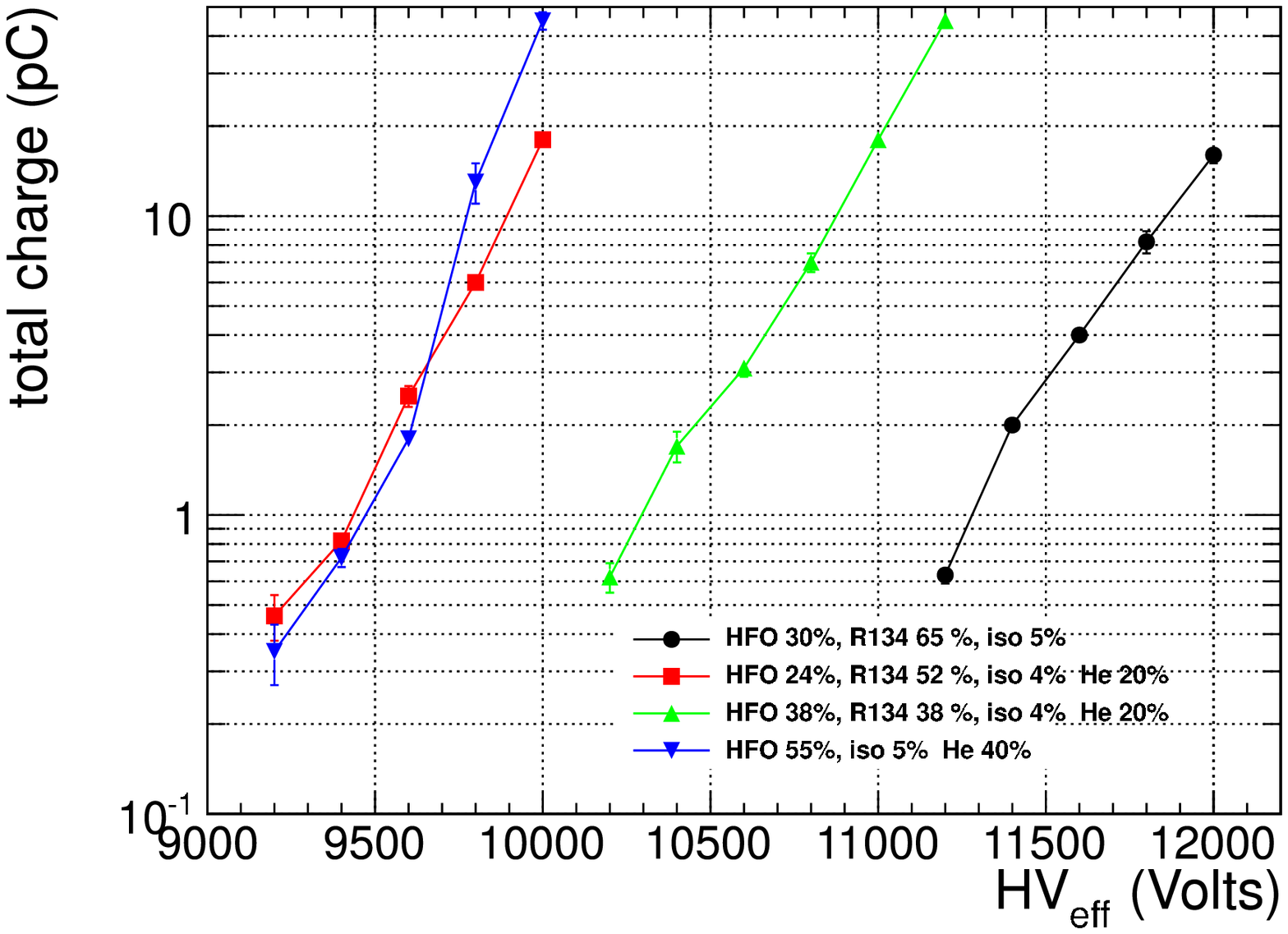}
    \caption{Integrated induced charge as a function of the effective voltage $HV_{eff}$,
for tetrafluoropropene/iso-buthane/Helium based gas mixtures}
    \label{fig:charge1}
  \end{center}
\end{figure}

This last mixture (shown with blue trianlges upside-down in the plot) is particularly interesting,
since it is an eco-friendly gas mixture (with a GWP<150), and the 2 mm RPC used in this
test were able to operate at a working voltage around 10 kV. 
Unfortunately, in this case, the range of voltage the RPCs could be operated is quite narrow,
since the streamer probability exceeds 40 $\%$ when efficiency is still below 90 $\%$.

\begin{figure}[tbp]
  \begin{center}
    \includegraphics[width=0.8\textwidth]{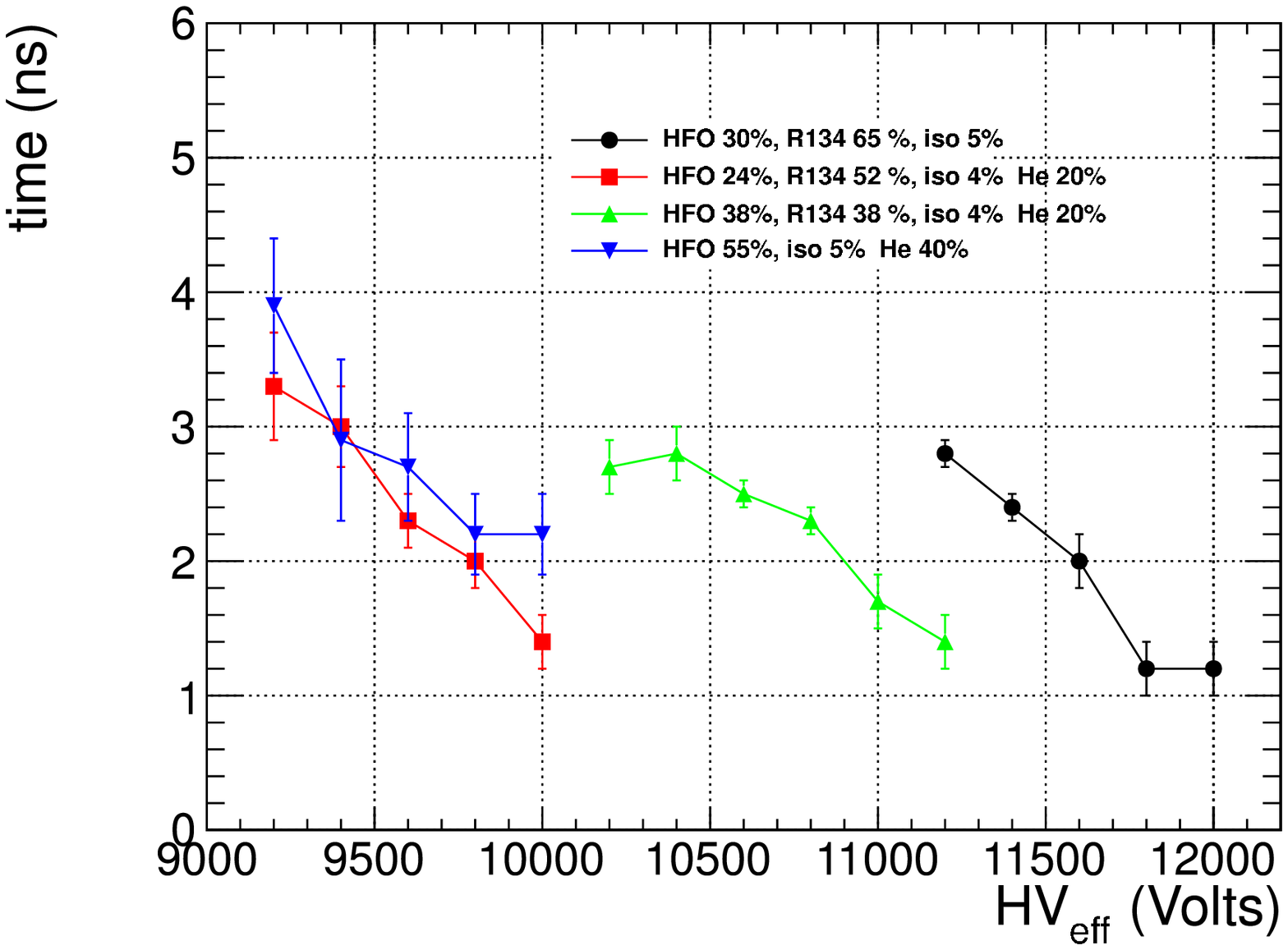}
    \caption{Time resolution as a function of the effective voltage $HV_{eff}$, 
for tetrafluoropropene/iso-buthane/Helium based gas mixtures} 
    \label{fig:time1}
  \end{center}
\end{figure}

The total induced charge and the time resolution as a function 
of the effective voltage $HV_{eff}$ for the gas mixtures tested are shown 
in Figures ~\ref{fig:charge1} and ~\ref{fig:time1} respectively. 
Due to the double pad readout used on the RPC used, the total induced charge 
should be divided by a factor about 2 in order to compare these results with 
devices in which the readout is done with a single pad or strips.

In order to reduce the streamer fraction with these gas mixtures, 
the effect of the addition of small quantities of SF$_{6}$ was studied. 
This is not a drawback, since SF$_{6}$ is used is so small quantities,
that the overall GWP of the gas mixture is still under the 150 limit value.

\begin{figure}[thbp]
  \begin{center}
    \includegraphics[width=0.8\textwidth]{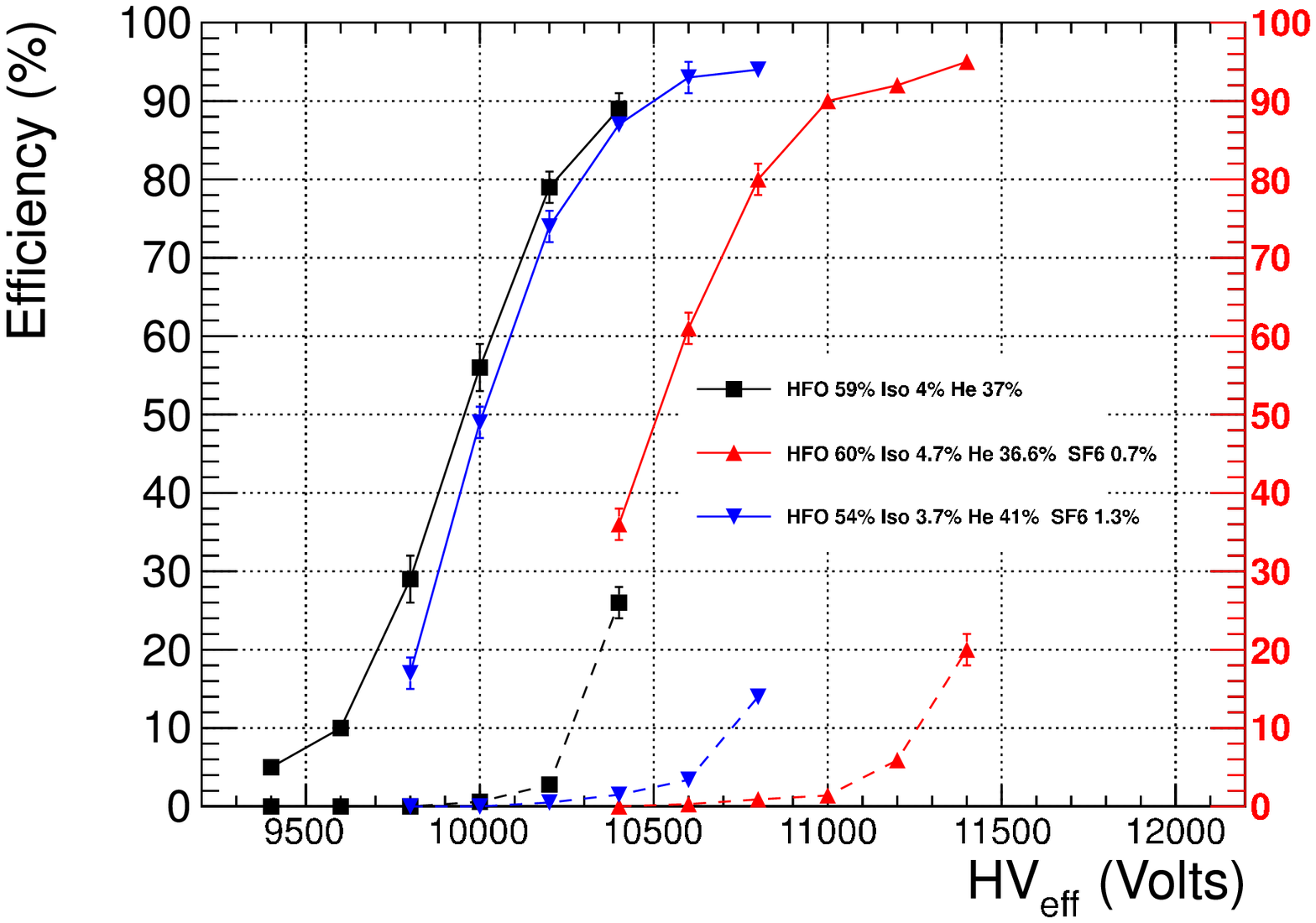}
    \caption{Efficiency (solid line) and streamer 
probability (dashed line) as a function of effective voltage $HV_{eff}$, 
for SF$_6$/tetrafluoropropene/iso-buthane/Helium based gas mixtures}
    \label{fig:eff2}
  \end{center}
\end{figure}

\begin{figure}[thbp]
  \begin{center}
    \includegraphics[width=0.8\textwidth]{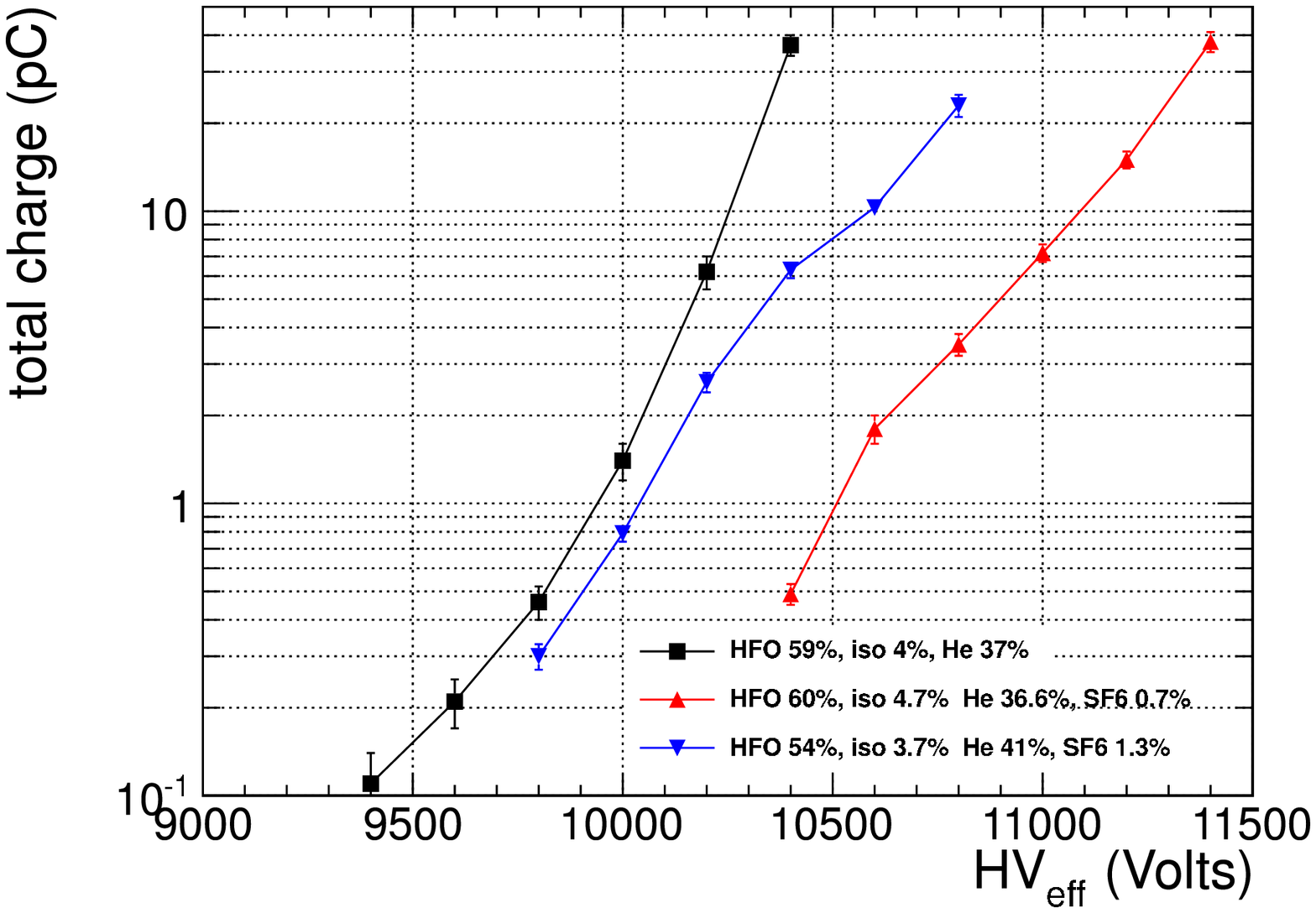}
    \caption{Integrated induced charge as a function of the 
effective voltage $HV_{eff}$, for SF$_6$/tetrafluoropropene/iso-buthane/Helium based gas mixtures}
    \label{fig:charge2}
  \end{center}
\end{figure}

\begin{figure}[thbp]
  \begin{center}
    \includegraphics[width=0.8\textwidth]{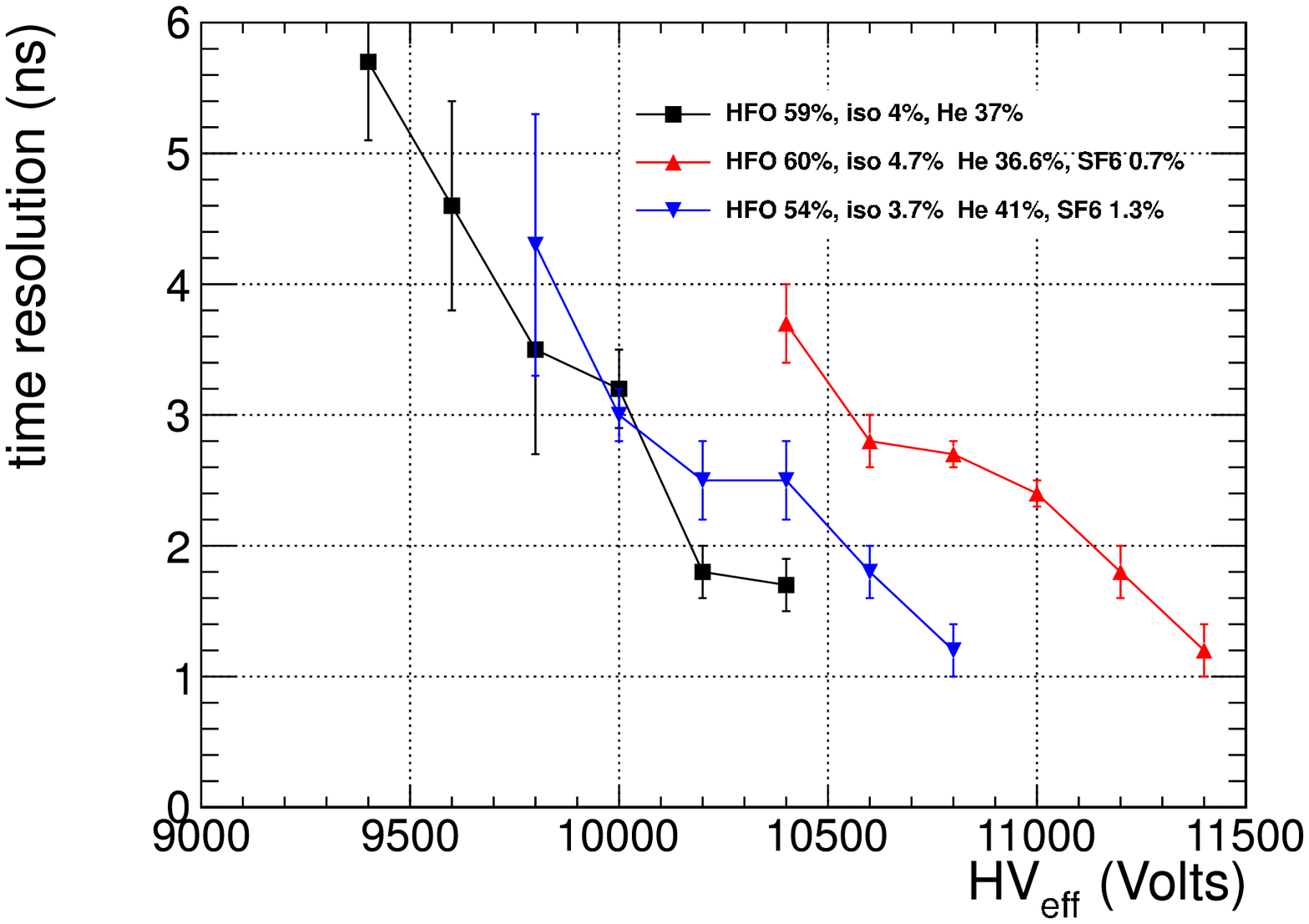}
    \caption{Time resolution as a function of the effective 
voltage, for SF$_6$/tetrafluoropropene/iso-buthane/Helium based gas mixtures}
    \label{fig:time2}
  \end{center}
\end{figure}

The efficiency, streamer probability, total charge and time 
resolution as a function of effective voltage 
for Helium, HFO-1234ze, iso-buthane and SF$_{6}$ based gas  mixtures
are shown in Figures ~\ref{fig:eff2}, ~\ref{fig:charge2} and ~\ref{fig:time2} respectively.
The presence of SF$_{6}$ fractions of the order of 1-1.5 $\%$ has the effect to limt 
the streamer probability to values below 20 $\%$ 
at efficienies greater than 90 $\%$ and represent a 
good starting point in the direction of an ecological gas mixtures 
to be used in the hostile environment of the LHC experiments.  

The performance of the gas mixtures tested are summarized in table ~\ref{tab:summary},
which reports the gas mixture, the effective voltage at 90 $\%$ efficiency (HV$_{90}$), 
the streamer probability, the total charge and the time resolution at HV$_{90}$.

\begin{table}[]
\centering
\caption{effective voltage at 90\% efficiency (HV$_{90}$), streamer probability, total charge
and time resolution at HV$_{90}$ for the gas mixtures investigated in this study}
\label{tab:summary}
\begin{tabular}{l|c|c|c|c|}
Gas mixture & HV$_{90}$ (kV) & Strm. prob. (\%) & ~2 Q$_{tot}$ (pC) & $\sigma_t$ (ns)\\ \hline
HFOze-iC$_4$H$_{10}$-R134a (30/5/65)               &  11.6   &   0.7   &  4  &   2   \\
HFOze-iC$_4$H$_{10}$-R134a-He (24/4/52/20)         &   9.8   &   1.2   &  6  &   2   \\
HFOze-iC$_4$H$_{10}$-R134a-He (38/4/38/20)         &  10.8   &   2     &  7  &   2.4 \\
HFOze-iC$_4$H$_{10}$-He (59/4/37)                  &  10.4   &  28     & 36  &   1.7 \\
HFOze-iC$_4$H$_{10}$-He-SF$_6$ (60/4.7/36.6/0.7)   &  11     &   1     &  7  &   2.4 \\
HFOze-iC$_4$H$_{10}$-He-SF$_6$ (54/3.7/41/1.3)     &  10.5   &   3     &  8  &   2.2 
\end{tabular}
\end{table}

\section{Conclusions}

The replacement of gases characterized by a Global Warming Power > 150 
in the gas mixtures presently used in RPCs is one
of the priorities in the R\&D on gaseous detectors in this period. 
Some of them could go out of production in the coming
years, while the community has to assure that the RPCs installed at 
the LHC experiments continue operating with a stable
performance up to the end of the High Luminosity 
phase of LHC, foreseen to be around 2030s.

The most natural replacement of tetrafluoroethane, the main 
component of the gas mixture standardly used in the 
RPC of the two main LHC experiments, is tetrafluoropropane, and 
many studies about its behavior are currently being
carried on.

In this paper, some tests have been presented, demonstrating 
the suitability of this gas as a possible component
of the future mixtures for RPCs, characterized by GWP < 150. 
However, its operating voltage is higher that 
tetrafluoroetane based gas mixtures, in particular higher 
than 10 kV, and this would require the use of specific
electrical components, with the corresponding additional costs. 

The solution could be represented by the addition of some 
fraction of Helium to these new gas mixtures, which
has the effect of reducing the operating voltage. Here the 
first tests using these new gas mixtures have been reported,
which demonstrate the effectiveness of this approach.

More tests are needed to find the optimal proportions these 
components have to be mixed, and this represents the plan
for the future.




\end{document}